\documentclass{article}
\usepackage{spconf,amsmath,graphicx}

\title{Sound Event Detection in Urban Audio With Single and Multi-Rate PCEN}
\twoauthors
  {Christopher Ick}
	{New York University\\
	Center for Data Science\\
	60 5th Ave., New York, NY 10011}
  {Brian McFee}
	{New York University\\
	Center for Data Science\\
	60 5th Ave., New York, NY 10011}

\begin{document}
%\ninept
%
\maketitle
\begin{abstract}
Recent literature has demonstrated that the use of per-channel energy normalization (PCEN), has significant performance improvements over traditional log-scaled mel-frequency spectrograms in acoustic sound event detection (SED) in a multi-class setting with overlapping events.
However, the configuration of PCEN's parameters is sensitive to the recording environment, the characteristics of the class of events of interest, and the presence of multiple overlapping events \cite{Lostanlen_2019}.
This leads to improvements on a class-by-class basis, but poor cross-class performance.
In this article, we experiment using PCEN spectrograms as an alternative method for SED in urban audio using the UrbanSED dataset, demonstrating per-class improvements based on parameter configuration.
Furthermore, we address cross-class performance with PCEN using a novel method, Multi-Rate PCEN (MRPCEN).
We demonstrate cross-class SED performance with MRPCEN, demonstrating improvements to cross-class performance compared to traditional single-rate PCEN.
\end{abstract}
\section{Introduction}
\label{sec:intro}

Noise suppression is a critical step in acoustic signal detection, particularly so in the case of practical sound event detection (SED) in field recordings.
Popular approaches to this task use convolutional operators, mimicking the methods used in computer vision \cite{8434391}.
When using audio as an input to these methods, the images are typically time-frequency representations (spectrograms).
Traditionally, log-scaling is the primary approach for noise reduction, and is the standard approach when spectrograms are the feature of interest.
For single-source audio in clean acoustic environments, this is sufficient for SED.

However, real world environments typically don't have clean, separated events of a single class; this is particularly true in the case of urban audio.
Field recordings often have multiple sound sources of varying acoustic qualities, leading to varying cross-class performance.
Furthermore, the introduction of auditory deformations can lead to rapid performance degradation \cite{DBLP:journals/corr/WangGHLS16}.

Former research has proposed using per-channel energy normalization (PCEN) as a time-frequency representation to mitigate the effects of background noise, demonstrating its use as an input to convolutional methods in SED.
PCEN has proven beneficial in various single-source tasks, including bioacoustic event detection \cite{Lostanlen_2019}, keyword spotting \cite{DBLP:journals/corr/WangGHLS16}, and vocal detection in music \cite{schlueter2018_ismir}.
The per-channel background suppression characteristics make PCEN an attractive choice in SED in urban environments, as background noise in urban environments tends to be Brownian, rather than Gaussian, an assumption made when log-scaling spectrograms \cite{8170052}.

One constraint of PCEN is its dependence on parameter configurations in relation to the specific acoustic properties of the sound event of interest \cite{8514023}.
This makes PCEN poorly suited for multi-class classification, as a given PCEN parameter configuration will likely suit only one particular class of sound events while performing poorly across other classes.
In this article, we demonstrate the effectiveness of PCEN for SED in urban environments.
Furthermore, we propose a new approach to acoustic event detection that combines the foreground separation characteristic of PCEN while preserving multi-class performance: Multi-Rate PCEN (MRPCEN).
By computing a multi-layered PCEN spectrogram at different parameter configurations, we gain the advantages of PCEN, particularly robustness to varying acoustic conditions and auditory deformations without the performance loss associated with cross-class performance.

\section{Audio Representation}
While log-scaling mel-spectrograms is a simple and computationally efficient method of range compression, it has limited effects in certain conditions, particularly those in which background noise is non-Gaussian noise.
PCEN has been used as a pre-processing method for time-frequency representations that reduces the effects of noise on convolutional neural networks by Gaussianizing the distribution of magnitudes across mel-frequency spectrogram coefficients \cite{DBLP:journals/corr/WangGHLS16}.
\begin{figure}
    \centering
    \includegraphics[width=\columnwidth]{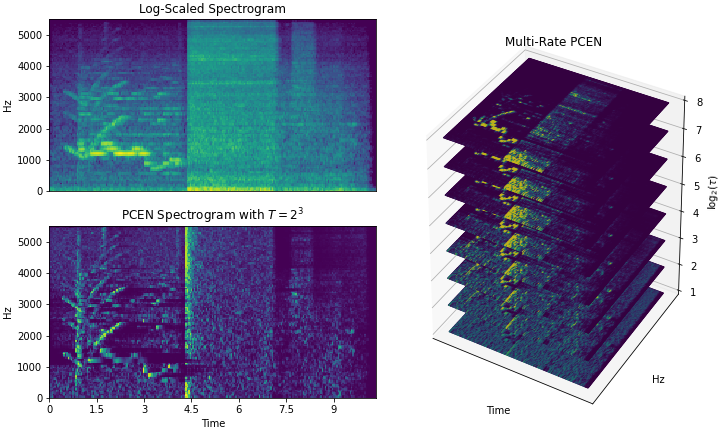}
    \caption{A comparison of standard SED audio featurizations compared to Multi-Rate PCEN. Note the unevenly distributed noise in the log-mel spectrogram, with more noise in the lower end of the spectrum (due to Brown noise found in urban environments), which is removed in the PCEN spectrogram.}
    \label{fig:pcen_spec_comparison}
\end{figure}

\subsection{PCEN}
PCEN is a sequence of audio processing steps on a spectrogram $\textbf{E}(t,f)$ using adaptive gain control scaled by the response an autogregressive filter $\phi_T$, followed by dynamic range compression. It takes the form:

\begin{equation}
    \textbf{PCEN}(t,f) = \left(\frac{\textbf{E}(t,f)}{(\epsilon + (\mathbf{E} * \phi_T)(t,f))^\alpha} + \delta \right)^r - \delta^r
\end{equation}

\noindent The gain $\alpha$ scales the smoothing of the spectrogram.
$\delta$ and $r$ are the bias and power of the dynamic range compression that PCEN provides.
We can see that the filter response $\mathbf{M}(t,f) = (\mathbf{E}*\phi_T)(t,f)$ scales our PCEN magnitude by:

\begin{equation}
    \mathbf{M}(t,f) = (\mathbf{E}*\phi_T)(t,f) = s \mathbf{E}(t,f) + (1-s)\mathbf{M}(t-\tau,f)
\end{equation}

\noindent where $0<s<1$ is the weight of 1\textsuperscript{st} order autogregressive filter (AR(1)), and $\tau$ is the discretized step defined by the hop size of the input spectrogram ($\epsilon$ is an offset term).
The resulting filter is a low-pass filter of 0dB gain and a cuttoff frequency of $\omega_c = \frac{2 \pi \tau}{T} = \arccos{(1-\frac{s^2}{2(1-s)})}$ at 3 dB, and sidelobe falloff of
10 dB per decade near $\omega_c$ \cite{8514023}.

We consider this filter response $\mathbf{M}(t,f)$ an approximation to the relative magnitude of stationary background noise at each frequency band $f$.
The effect of scaling our output PCEN spectrogram by the reciprocal of this response is amplifying/magnifying the response of foreground events and suppressing background events on a per-channel basis.
In regimes where the background noise isn't Gaussian, such as the case with Brownian noise in urban environments, this decorrelates noise across different frequency bands.

A critical tuning parameter of this smoothing of the autoregressive filter is the rate parameter $T$, which defines the cutoff frequency by $\omega_c = \frac{2 \pi \tau}{T}$.
Setting $T$ too low will reduce the noise reduction effects of PCEN, but setting it too high will suppress the sound event of interest, especially if the sound event of interest is stationary.
Prior practical recommendations for the rate parameter are defined by the stationarity of the sound, the frequency range, and the chirp rate.
For single sources in a consistent acoustic environment, tuning the rate parameter $T$ to these specifications proves sufficient for audio detection and classification tasks \cite{8514023}.

\subsection{Multi-rate PCEN}
In a field setting, with several audio classes with varying acoustic properties and variable recording conditions, a singular ideal value for $T$ is much harder to identify.
For example, in the case of urban audio, the combination of short, fast decay sounds (such as gunshots or dog barks) compared to longer ambient sounds (e.g. sirens, air conditioners) lead to widely varying preferred values of $T$.
To capture information across varying regimes of our rate parameter, $T$, we take inspiration from 3-channel RGB images in computer vision.
These three separate but correlated feature maps can be passed to a convolutional neural network (CNN) which can use this multi-frequency image to make predictions that a black-and-white image could not.

We replicate this multi-regime approach, but instead of frequency responses, we vary our rate parameter $T$ in each layer, as shown in figure \ref{fig:pcen_spec_comparison}.
Each $i$\textsubscript{th} layer of the image has a differing level of gain control applied based on the cutoff threshold from $T_i$.
By using multiple logarithmically-scaled values of $T_i$, we can produce a multi-layered image that captures information of sounds at varying decay lengths.
This ensures that sound events that may be suppressed using one value of $T_i$ will be preserved in another.
The resulting multi-channel image can be used as an input to a CNN much like a multi-channel color image is used in machine vision.
By incorporating information at varying degrees of gain control, our model preserves the robustness of PCEN without degrading multi-class performance.

\section{Experimental Design}

\subsection{UrbanSED}
We used the UrbanSED dataset to evaluate the performance on this method for the task of sound event detection\cite{8170052}.
UrbanSED contains 10,000 synthetic soundscapes of 10 distinct sound classes, with each class having approximately 1000 instances per class, drawn from the UrbanSound8K dataset.
Each soundscape contains 1-9 time/class labeled foreground events with additive background Brownian noise.
The dataset is pre-split in a 6-2-2 training, validation, and evaluation subsets.
Because UrbanSED is synthetic, we can ensure no spurious unlabeled audio is included in the soundscapes, making it a standard benchmark for state-of-the-art SED models.

\subsection{Data augmentation}
Because one of the primary predicted benefits of PCEN is improved robustness to audio deformations, we augmented our dataset with several reverberant duplicates of UrbanSED.

In this implementation, reverb is modeled as a convolution of a source signal and an impulse response of a given acoustic environment.
Augmenting a given audio clip with reverb is done by computing the convolution of the impulse and the audio file.
This effectively computes the response of the audio clip in the acoustic environment associated with the impulse.
We used 6 distinct reverb responses, three impulse responses recorded in different acoustic environments, and three synthetic impulse responses of white noise with an exponential decay envelope $e^{-t/\tau_c}$. 

The three real impulse responses were recorded in a bedroom\footnote{``My Bedroom'', https://freesound.org/people/Uzbazur/sounds/382907/}, an alleyway\footnote{``alley.wav'', https://freesound.org/people/NoiseCollector/sounds/126804/}, and a tunnel\footnote{``tunnel\_2013'', https://freesound.org/people/recordinghopkins/sounds/175358/}, each with increasing decay times.
The three synthetic impulse responses had decay time constants $\tau_c$ of $0.1$s, $0.3$s, and $0.5$s.
In addition to reverb augmentation, we also duplicated our dataset by pitch shifting each sample by \{$\pm 1, \pm 2$\} semitones.
We applied pitch shifts and convolutional to reverb to our dataset using using \textit{MuDA}, a library for musical data augmentation \cite{muda2015}.

\subsection{Audio featurization}
Training audio was processed using librosa 0.7.2 \cite{brian_mcfee-proc-scipy-2015} which generated a PCEN spectrogram for each audio sample and rate parameter $T$.
Each spectrogram had the following parameters: sampling rate $44.1$kHz, window size $1024$ samples, hop length $512$ samples, and $128$ mel-frequency bands. 
Our resulting frame rate was $44.1$kHz$/512 \approx 86$Hz.
At a 10 second length sample length, each spectrogram had a horizontal length of 862 samples per band.

Our rate parameters were 10 logarithmically spaced values, ranging from $2^0=1$ to $2^9=512$, corresponding to averaging over windows ranging from $10$ms to $6$s.
The adaptive gain control bias was set to $\epsilon=10^{-6}$, the gain was set to $\alpha = 0.98$, the dynamic range compression bias was set to $\delta=2$, and the compression power $r=0.5$ for all rate configurations; these parameters were chosen based on prior work in audio classification \cite{Lostanlen_2019}.
Each audio sample had 10 PCEN spectrograms, one per rate parameter.
Individual layers/subsets were selected in training and evaluation.
Following the channels-last conventions of Keras, the resulting PCEN spectrograms have the shape of (128,862,10), and the resulting log-mel spectrograms had a shape (128,862,1).

\subsection{Network architecture}
The network architecture is inspired by the $L^3$ audio sub-network \cite{DBLP:journals/corr/ArandjelovicZ17} for discriminative audio-video correspondence embeddings.
This architecture has demonstrated success in its use for classification and predictions at a fine time-resolution with mel-spectrogram inputs\cite{8434391}.
We follow the implementation of this architecture found in \cite{8434391}, with the input layer adjusted to accept multi-layer images to ensure that all methods were being evaluated on near-identical architecture.
Additional parameters in the input layer may lead to lower performance due to additional depth requirements, but this is assumed to be negligible for this application.

Models were built and trained using Keras 2.3.1 \cite{chollet2015keras}, with Tensorflow 2.2.0 \cite{tensorflow2015-whitepaper}.
The model was trained using the Adam optimizer \cite{kingma2014method}, using UrbanSED's pre-folded training and validation sets.
The loss function was binary cross-entropy on a per-class per-frame bases.
The validation metric was accuracy.
If no improvement in the accuracy was seen in 10 epochs, the learning rate would be reduced.
Early stopping was implemented if no improvement was seen after 30 epochs.
Evaluation was handled via the sed\_eval package \cite{sed_eval_art}, which computed segment-based classification metrics, including precision, recall, and error rate.
We used the F1-score, the harmonic mean of precision and recall, as our main metric of effectiveness.
These metrics were computer both per-class and overall.
For reproducability, the implementation and experimental framework, including data preprocessing, model training, and evaluation, is publicly available on github\footnote{https://github.com/ChrisIck/pcen-t-varying}.

\subsection{Data preparation}
We trained and evaluated models across multiple datasets to test robustness and stability to audio deformation.
We primarily used the \textit{dry} dataset as our baseline dataset, which was UrbanSED with no reverb augmentation, to ensure we were achieving state of the art performance seen in \cite{8434391}.
The \textit{realreverb} set is built from the reverb-augmented audio using the 3 real recorded impulse responses, and is the primary set we evaluate, as this is closest to real-world conditions.
The \textit{simreverb} set, using the remaining 3 synthetic impulse responses, demonstrated the strongest separation between PCEN and log-mel models, but the results don't generalize as well as when using real impulse responses.
All models were trained on data augmented by $\{\pm 1, \pm 2\}$ pitch-shifts, but were validated and evaluated on non-shifted data.

\subsection{Rate parameters}
We trained and evaluated models on fixed sets of rate parameters.
We experimented with single rate parameter models and varying sets of multi-layered models, each with rate constants in $T_k = 2^k$ for $k \in \{0,1,...,9\}$.
We tested a total of 44 unique rate parameter configurations for our PCEN models.
This included single-rate parameter models and $n$-layer models including rate parameters from $T_i$ to $T_{i+n-1}$.
As a baseline, we also computed a model using a traditional log-scaled mel-spectrogram, as seen in previous literature for this application and dataset \cite{8434391}.

\subsection{Training and evaluation}
Each set of models was trained and evaluated on both datasets to see how stable each model was to reverberation conditions that were both contained in the respective training dataset, as well as in conditions distinct from those in the training set.
Evaluation was done on bootstrapped subsamples of each evalution set, sampling 100 evaluation examples with replacement, 100 times per model and dataset.
Evaluations were completed both in overall micro-averaged performance metrics across classes, as well as on a per-class basis.

\section{Results}
\begin{figure}
    \centering
    \includegraphics[width=\columnwidth]{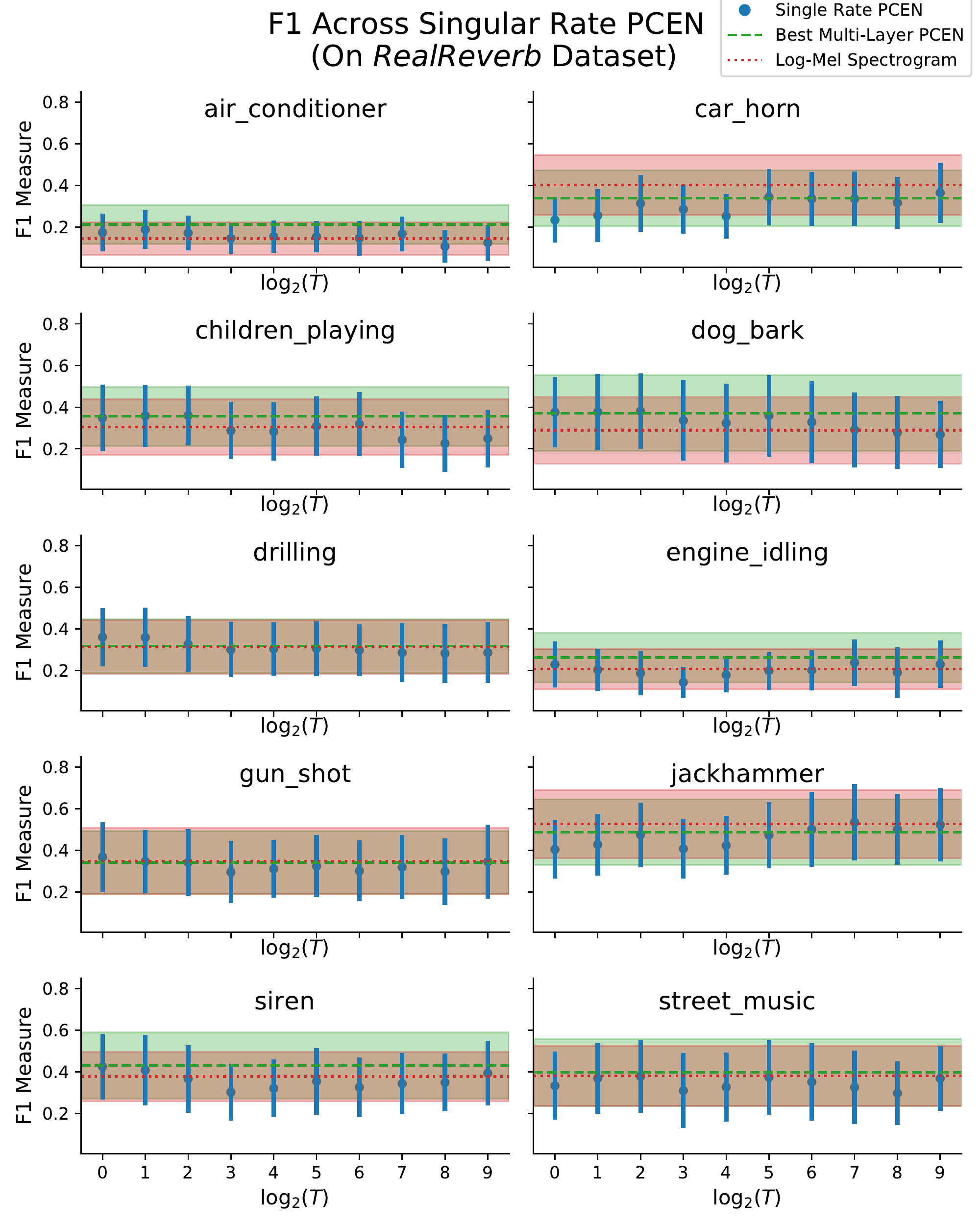}
    \caption{F1-score of single-rate PCEN models at various $T$ values (blue dots), compared to 8-layer MRPCEN (dashed green) and log-mel (dotted red) models evaluated on the 100 bootstrapped samples of the \textit{realreverb} dataset.}
    \label{fig:singlerate_compare}
\end{figure}

\subsection{Per-class performance}
We demonstrate PCEN's robustness to noise by comparing performance of models trained and evaluated on the \textit{realreverb} dataset.
We see in figure \ref{fig:singlerate_compare} that single rate PCEN demonstrates performance at or above log-scaled spectrograms, depending on its choice of rate parameter $T$ and the class of the target event.
Certain classes, like the jackhammer and gunshot classes, prefer higher $T$ values in the neighborhood of $T = 2^7$ where mid-low frequency is filtered out with the adaptive gain aspect of PCEN.
Similarly, for higher and full-band frequency stationary events, such as air conditioners and sirens, lower values of $T$ in the neighborhood of $T \in [2^1,2^3]$ perform better, as the low-frequency noise doesn't interfere with the relevant frequency bands.
However, it is crucial to note that due to the diversity of sonic characteristics of these different classes, there does not exist a single ideal value of $T$ that produces the strongest performance overall.

\subsection{Cross-class performance}
MRPCEN successfully outperforms most single-rate models in cross-class performance, as seen in figure \ref{fig:singlerate_compare}.
The plot shows per-class performance of each model evaluated on the \textit{realreverb} dataset, compared to a log-mel based model and an 8-layer MRPCEN.
In comparison, MRPCEN model preserves information at multiple rates, it consistently performs at or above the majority of the single rate models.
Furthermore, MRPCEN performs at or above the level of the standard log-mel spectrogram with the exception of the ``car\_horn'' and ``jackhammer'' classes, providing it with the highest overall performance metric on this dataset (compared to the strongest performing single-rate PCEN model $T=2^1$) in table \ref{fig:overall_f1_table}.

\begin{table}
    \centering
        \begin{tabular}{l|p{1.7cm}|p{1.7cm}|p{1.7cm}}
        Model & Overall F1 (\textit{realreverb}) & Overall F1 (\textit{simreverb}) & Overall F1 (All Data)\\
        \hline
        Logmel & 0.334 & 0.175 & 0.274\\
        PCEN & 0.345 & 0.167 & 0.268\\
        MRPCEN & \textbf{0.356} & \textbf{0.204} & \textbf{0.285}
    \end{tabular}
    \caption{Overall Micro-averaged F1 Scores on the \textit{realreverb} and \textit{simreverb} datasets}
    \label{fig:overall_f1_table}
\end{table}

\section{Conclusion}
The results here show PCEN as a viable alternative to log-mel spectrograms, showing equivalent or improved performance depending on rate parameter choice, which in turn depends on the acoustic characteristics of the target sound event and acoustic environment.
We can also see that MRPCEN provides cross-class performance improvements over single-rate PCEN models.
MRPCEN permits simultaneous prediction of differing classes with distinct acoustic characteristics in a single model.
In a field setting, this will lead to less per-class knowledge and tuning expertise needed to effectively deploy a model that can perform well in multi-class applications.

\vfill\pagebreak
\section{Acknowledgements}
This work was supported in part through by NSF award 1955357, and in part by the NYU IT High Performance Computing resources, services, and staff expertise.

%\section{REFERENCES}
%\label{sec:refs}

% References should be produced using the bibtex program from suitable
% BiBTeX files (here: strings, refs, manuals). The IEEEbib.bst bibliography
% style file from IEEE produces unsorted bibliography list.
% -------------------------------------------------------------------------
\bibliographystyle{IEEEbib}
\bibliography{strings,refs}

\end{document}